\documentclass[preprint,showpacs,preprintnumbers,amsmath,amssymb]{revtex4-1}
\usepackage[english]{babel}
\usepackage{graphicx}
\usepackage{xspace}
\usepackage{fancybox}
\usepackage{epic}
\usepackage{calc}
\usepackage{ifthen}
\usepackage{float}
\usepackage{color}
\usepackage{amsmath,amssymb}
\usepackage{rotating}
\usepackage{subfigure}
\usepackage{exscale}
\usepackage{setspace}
\usepackage[latin1]{inputenc}
\begin{document}
\title{Physical properties and band structure of reactive molecular beam epitaxy grown oxygen engineered HfO$_{2\pm x}$}
\author{Erwin Hildebrandt}\email{emh@oxide.tu-darmstadt.de}
\affiliation{Institute of Materials Science, Technische
Universität Darmstadt, 64287 Darmstadt, Germany}
\author{Jose Kurian}
\affiliation{Institute of Materials Science, Technische
Universität Darmstadt, 64287 Darmstadt, Germany}
\author{Lambert Alff}\email{alff@oxide.tu-darmstadt.de}
\affiliation{Institute of Materials Science, Technische
Universität Darmstadt, 64287 Darmstadt, Germany}

\date{25 July 2012}

\begin{abstract}

We have conducted a detailed thin film growth structure of oxygen engineered monoclinic HfO$_{2\pm x}$ grown by reactive molecular beam epitaxy (MBE). The oxidation conditions induce a switching between ($\overline{1}11$) and ($002$) texture of hafnium oxide. The band gap of oxygen deficient hafnia decreases with increasing amount of oxygen vacancies by more than 1 eV. For high oxygen vacancy concentrations, defect bands form inside the band gap that induce optical transitions and $p$-type conductivity. The resistivity changes by several orders of magnitude as a function of oxidation conditions. Oxygen vacancies do not give rise to ferromagnetic behavior.

\end{abstract}

\maketitle


\section{Introduction}

Within the past years, intense research has been carried out on hafnium dioxide which is among the best candidates to fully replace SiO$_2$ as gate dielectric for microelectronics.\cite{Choi:11,Wilk:01} Replacing the established Si/SiO$_2$ system in complementary metal oxide semiconductor (CMOS) technology raised various challenges to be solved, especially process compatibility with existing CMOS fabrication methods, to facilitate the usage of existing production lines, capacities and know-how. Among these challenges are maximizing the dielectric constant, electronic structure alignment, substrate and contact material compatibility (chemical, structural, and electrical), and control of formation of unwanted interfacial layers, just to name a few. A variety of deposition methods has been utilized in the growth of hafnium oxide thin films: atomic layer deposition,\cite{Niinisto:09,Kukli:05} sputtering,\cite{Kuo:92} pulsed laser deposition,\cite{Wang:08} chemical vapour deposition,\cite{Ohshita:01} e-gun evaporation,\cite{Lehan:91} high-energy ion beam assisted deposition,\cite{Mori:03} and sol-gel methods.\cite{Nishide:00,Neumayer:01} All mentioned deposition methods do have unique features and certain advantages and disadvantages for the growth of HfO$_2$ thin films. As an example, PLD as a simple and flexible deposition technique allows large deposition rates but lacks sufficient stoichiometry control.\cite{Choi:11}

Unexpected physical properties of hafnium oxide thin films have been revealed which could qualify hafnia for various additional applications in microelectronics beyond CMOS. The report of room temperature ferromagnetism or $d^0$-ferromagnetism, which denotes ferromagnetism in transition metal oxides with closed shell configuration, shows that HfO$_2$ could also be a candidate for applications in spintronics.\cite{Venkatesan:04} The most recently discovered phenomenon in hafnia is a resistive switching effect as a function of applied voltage for application in non-volatile memories.\cite{Lee:07,Walczyk:11,Walczyk:09} Both phenomena, resistive switching and $d^0$-/room temperature ferromagnetism would significantly boost the importance of hafnium oxide for microelectronics, since it has already been qualified for standard CMOS processing.
All applications of HfO$_2$ thin films do require a precise and detailed understanding of film growth, structure and physical properties as a function of deposition parameters, and in particular as a function of stoichiometry. Control of thin film stoichiometry is a key to control the physical properties of hafnia. The defect or oxygen vacancy formation has been modeled theoretically and identified to be crucial. In the case of $d^0$-ferromagnetism oxygen vacancy induced charge carriers are responsible for the magnetic coupling.\cite{Venkatesan:04,Coey:05} In the case of resistive switching, the local change of oxygen defects may lead to conducting paths through the film. For the classical application as a gate dielectric, the maximization of dielectric constant, $\kappa$, has been investigated by, e.g., stabilizing tetragonal ($t$-HfO2) or cubic ($c$-HfO2) phases of HfO$_2$ exhibiting higher $\kappa$-values than the monoclinic ($m$-HfO2) structure. The stabilization of different phases was expected to be supported by substitutions with, e.g., Si, C, Ge, Sn, Ti, and Ce.\cite{Dutta:09,Fischer:08} All the above mentioned properties (high-$\kappa$ dielectric, $d^0$-/room temperature ferromagnetism, resistive switching) controllable in one single material by oxygen engineering, would turn hafnium oxide into a real multifunctional material with great future prospective.

In this study, reactive molecular beam epitaxy (RMBE) has been used to grow HfO$_{2\pm x}$ thin films with $x$ varying over a wide range. Since RMBE allows to separately control the hafnium atomic flux and the oxidation conditions, and additionally allows to vary other deposition parameters with high flexibility, it is an ideal tool to vary/control film stoichiometry and the formation of oxygen vacancies. In order to investigate the impact of stoichiometry in HfO$_{2\pm x}$, various films with thicknesses between 50 and 200 nm have been grown under different oxidation conditions and have been carefully investigated regarding their structural, optical, electrical, and magnetic properties.


\section{Experimental}

The RMBE unit used for this study consists of a custom made ultra-high vacuum (UHV) chamber (base pressure $\sim10^{-9}$ mbar) with a load lock arrangement.\cite{Alff:11} High-purity metal hafnium (99.9\%, MaTecK) was evaporated using e-beam evaporation, and {\em in situ} oxidation was achieved by means of oxygen radicals supplied by rf-activated oxygen (purity 99.995\%). Substrates were radiantly heated by a plate resistance heater with a temperature capability of more than 900 $^\circ$C. For {\em in situ} rate monitoring and control, quartz crystal microbalances (QCMs) are connected to a state-of-the-art four channel deposition controller (Cygnus, Inficon). The key deposition parameters varied in this study are substrate temperature ($T_{\text{S}}$, varied between $400-830$ $^\circ$C), oxygen flow rate (varied between 0 and 2.5 sccm), and rf-power applied to the radical source (varied between 0 and 300 W). The structural characterization was carried out {\em in situ} using reflection high-energy electron diffraction (RHEED) and {\em ex situ} using an X-ray diffractometer with Cu K$_\alpha$-radiation (Rigaku SmartLab, 9 kW). A magnetic property measurement system (MPMS, Quantum Design) was used for obtaining magnetization data as a function of temperature and applied field. A custom made four-probe measurement setup was used for measuring resistivity as a function of temperature. The Hall coefficient was obtained utilising a custom made setup with van-der-Pauw geometry in the MPMS cryostat. For optical characterization, a photospectrometer (PERKIN ELMER Lambda 900) was used in a wavelength range from 175-2000 nm in transmission geometry. Surface topography was investigated by atomic force microscopy (MFP-3D Stand Alone AFM, Asylum Research).

Since the focus of this work is to study the effect of oxygen stoichiometry in hafnium oxide thin films on its structural, optical, and electrical properties rather than focusing on CMOS process compatibility, $c$-cut and $r$-cut sapphire substrates were chosen instead of silicon wafers. Substrate sizes have been 7 x 4 and 5 x 5 mm$^2$ single side polished and 7 x 7 mm$^2$ double side polished for optical characterization. Sapphire substrates are chemically more inert compared to silicon substrates, which allows higher deposition temperatures necessary to obtain highly crystalline films. Chemical reaction between substrate and film, the formation of unwanted mixed phases at elevated temperatures, and the formation of interfacial layers are minimized. This allows to study the effect of oxygen deficiency on thin film material properties while ruling out substrate contributions as much as possible. Besides, sapphire provides clean crystal surfaces for potential epitaxial growth of HfO$_2$ without the necessity to apply SiO$_2$-stripping procederes. The substrates used for this study were thoroughly cleaned before use by treatments with H$_2$SO$_4$ and H$_2$O$_2$ mixtures (H$_2$SO$_4$ : H$_2$O$_2$ = 1 : 1) in an ultrasonic cleaner. After chemical cleaning, substrates were annealed in air at 1050 $^\circ$C for 12 h in order to obtain high surface quality.\cite{Yoshimoto:95} In all deposition runs two sapphire substrate orientations, $c$- and $r$-cut, were used. Section III is mostly limited to the description of the results obtained for $c$-cut sapphire. We have restricted the detailed analysis to films on $c$-cut sapphire substrates due to their relatively higher crystallinity. This means that for $r$-cut substrates different growth parameters have to be chosen. However, as far as we can judge from our limited analysis of the thin films on $r$-cut substrates films, similar trends are observable.


\section{Results and Discussion}

\subsection{Thin film growth and structural analysis}

We first describe the thin film growth of HfO$_{2-x}$ by investigating separately the influence of the key parameters substrate temperature, oxygen flow rate through the radical source, and rf-power applied to the radical source on the structural properties crystallinity, phase purity, stoichiometry, and surface topography. The Hf metal evaporation rate was held constant at 0.7 {\AA}/s, which is a moderate and well controllable rate for Hf evaporation via e-gun.


\begin{figure}[t]
\centering{\includegraphics[width=0.9\columnwidth,bbllx=0,bblly=0,bburx=574,bbury=412,clip=]{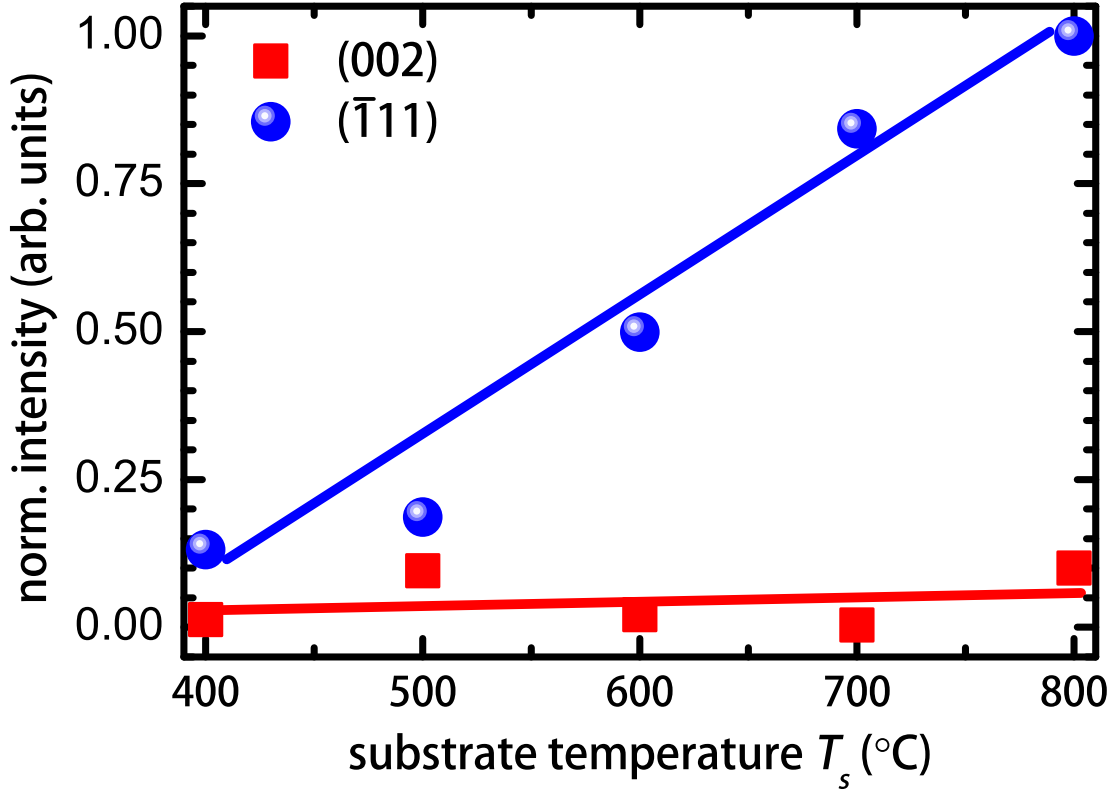}}
\caption{(Color online) Normalized intensities of ($\overline{1}11$) and ($002$) reflections of 50 nm thick hafnia thin films as a function of substrate temperature (Hf-rate 0.7 {\AA}/s, 1.5 sccm oxygen flow rate, 200 W rf-power). Solid lines are guides to the eye.}\label{Fig:Ts}
\end{figure}

\subsubsection{Deposition temperature}

A series of 50 nm thick hafnium oxide thin films has been grown on $c$-cut sapphire substrates. The substrate temperature was varied from 400 to 800 $^\circ$C in steps of 100 $^\circ$C while keeping other deposition parameters constant (rf-power applied to the radical source (200 W), and oxygen flow rate (1.5 sccm)). Note that the indicated temperature values are set temperatures, the real temperature may differ by about 5\%. All plotted data has been obtained within one continuous growth series. In all cases, hafnium oxide stabilized in the monoclinic phase with high texture but with different thin film orientation with respect to the substrate. We have quantified this change of orientation by monitoring the intensity of the strongest reflections of $m$-HfO$_2$, ($\overline{1}11$) and ($002$), as a function of substrate temperature as shown in Fig.~\ref{Fig:Ts}. The higher the substrate temperature during growth, the higher becomes the intensity of the ($\overline{1}11$) reflection, whereas the intensity of the (002) reflection does not change systematically. Around 700 $^\circ$C the intensity of the ($002$) peak is almost vanishing leading to nearly single oriented, highly textured HfO$_2$ films. At higher and lower temperatures, but particularly at 500 $^\circ$C, the ($00l$) orientation is reoccurring. No significant shift in the 2$\theta$ angle as a function of substrate temperature of both, ($\overline{1}11$) and ($002$) reflection, could be observed. The rocking curves of both reflections have values of the full width at half maximum (FWHM) in the range of 8$^\circ$, in-plane measurements reveal disorder in the $a$-$b$-plane. The increasing intensity and decreasing FWHM of the ($\overline{1}11$) reflections with substrate temperature shows, that increasing temperature leads to increased crystallinity, crystal size, and preferred ($\overline{1}11$) orientation, which is in good agreement with findings from Bharathi {\em et al}.\cite{Bharathi:10} As reported in literature, for growth temperatures below 400 $^\circ$C HfO$_2$ usually is amorphous, whereas at higher temperatures crystallites start to form leading to mixed films of amorphous and polycrystalline fractions.\cite{Aarik:99,Ho:03,Cheynet:07} In the case of bulk HfO$_2$, a transition from the monoclinic phase to the tetragonal phase is reported for elevated temperatures, \cite{Ruh:70} which could not be observed for our thin films up to 800 $^\circ$C.
No reflections of other crystallographic modifications such as orthorhombic ($o$-HfO$_2$) and $c$-HfO$_2$ could be observed.\cite{Wang:92,Adams:91,Benyagoub:03} We have not investigated the film growth at higher substrate temperatures, since film-substrate interactions leading to mixed compounds of Al$_x$Hf$_y$O$_z$ are expected. Also, oxygen exchange with the substrate may occur as reported for Si-substrates at much lower temperatures.\cite{Goncharova:06}


\begin{figure}[t]
\centering{%
\includegraphics[width=0.9\columnwidth,bbllx=0,bblly=0,bburx=576,bbury=410,clip=]{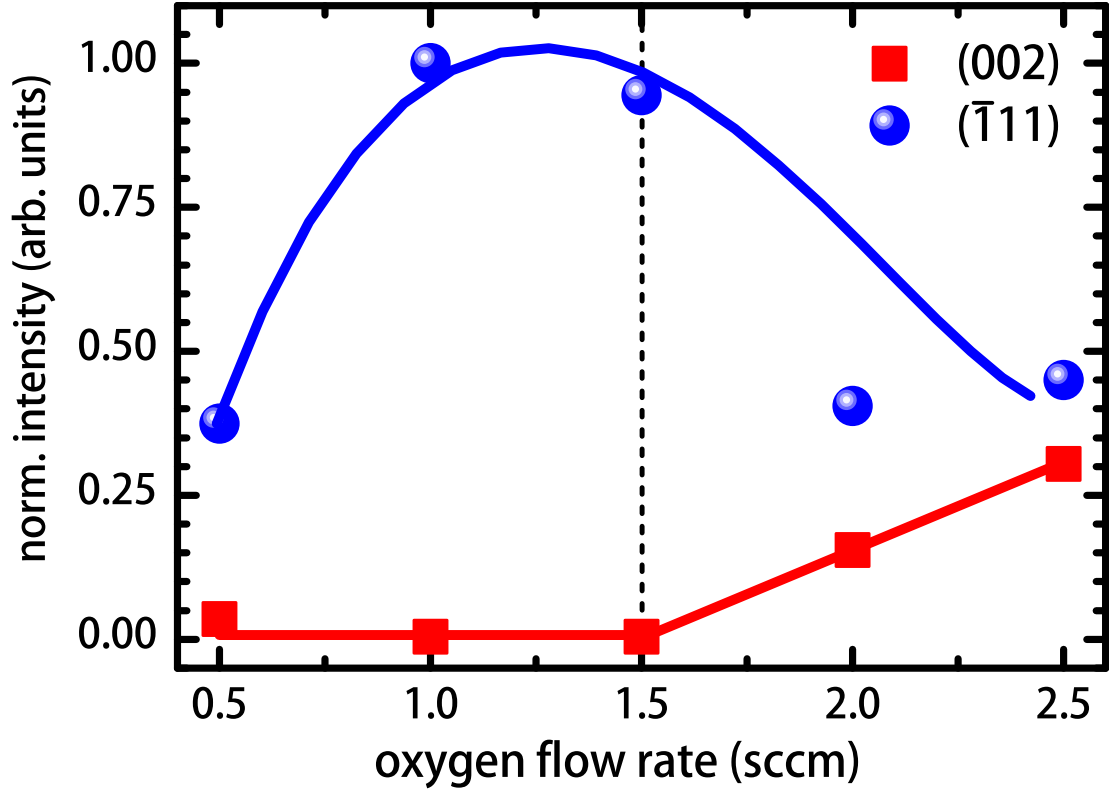}}
\caption{(Color online) Normalized intensities of ($\overline{1}11$) and ($002$) reflections of 50 nm thick hafnia thin films as a function of oxygen flow rate (Hf-rate 0.7 {\AA}/s, $T_{\text{S}}$ = 700 $^\circ$C, 200 W rf-power), solid lines are guides to the eye. The vertical dashed line indicates approximately the oxygen flow rate corresponding to stoichiometric composition.}\label{Fig:Oflow}
\end{figure}

\subsubsection{Oxygen flow rate}

The oxygen flow rate was varied between 0.5 and 2.5 sccm of molecular oxygen supplied to the radical source, covering a wide range of oxidation conditions during growth. This allows the controlled introduction of oxygen deficiencies in HfO$_{2-x}$ and over-oxidation in HfO$_{2+x}$ leading to oxygen interstitials and/or hafnium vacancies. As already observed for the variation of substrate temperature, at all oxygen flow rates hafnia stabilized in its monoclinic phase, highly textured, but with two orientations, ($\overline{1}11$) and ($002$). Fig.~\ref{Fig:Oflow} shows the dependence of the normalized intensities of both orientations as a function of oxygen flow rate. The intensity of the ($\overline{1}11$) reflection is highest in the range of 1.0 to 1.5 sccm, where the (002) reflection vanishes.
For low oxygen flow, the clearly dominanting orientation is ($\overline{1}11$). For higher flows, there is an increasing amount of ($002$) orientation. It is remarkable, that although we have varied the oxygen content in hafnia in a wide range, the monoclinic structure sustains and does not collapse, neither due to oxygen deficiencies nor due to excess oxygen. Even when reducing the oxygen content further (flow rates below 0.5 sccm), neither metallic hafnium could be observed, nor a collapse of the crystal structure of $m$-HfO$_2$ occurred. Metallic hafnium started to form only when molecular oxygen with a much lower oxidation potential was supplied during growth at a flow rate of about 0.2 sccm (details are described below). No traces of $c$-HfO$_2$, $o$-HfO$_2$, or $t$-HfO$_2$ could be identified. Oxygen flow rate studies on the growth of hafnium oxide thin films and their influence on film orientation have been carried out earlier, but were more focussed on device properties than on crystallinity and film orientation.\cite{Cherkaoui:08,Moon:05}


\begin{figure}[t]
\centering{%
\includegraphics[width=0.9\columnwidth,bbllx=0,bblly=0,bburx=224,bbury=88,clip=]{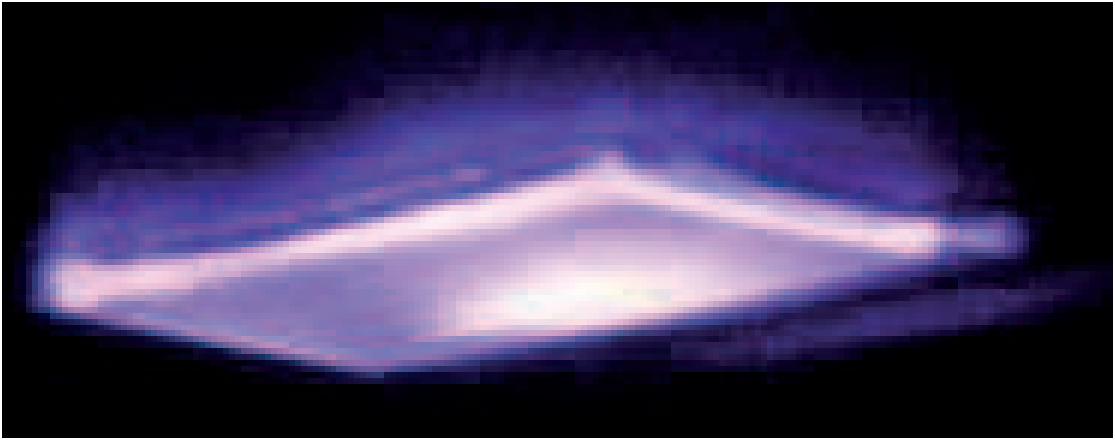}}
\caption{(Color online) Exemplary image of cathodoluminescence of a reduced HfO$_{2-x}$ thin film on $r$-cut sapphire irradiated by 30 keV electrons from a RHEED gun. The same effect is also observed for thin films on $c$-cut substrates.}\label{Fig:Nice}
\end{figure}

\subsubsection{RF power applied to the radical source}

The higher the applied rf power, the stronger the plasma intensity monitored via photocell and the higher the cracking efficiency at a given oxygen flow rate. The rf power was varied between 0 and 300 W. At 300 W, very high plasma intensities and, thus, high oxidation conditions are obtained. At such high rf power, the X-ray intensities of the ($\overline{1}11$) reflection are highest, however, rf back reflection is strongly increased at 300 W. Therefore, we decided to work in the range of 200 W. Between 100 W and 200 W no measurable influence of the rf power on the X-ray intensity could be observed. We have worked with the growth parameters where the formation of the ($002$) orientation is completely suppressed. The change in rf power did not lead to a reentrance of the ($002$) orientation.

For 0 W rf-power, meaning that no radicals are created and, thus, only O$_2$ molecules were provided for oxidation, the normalized ($\overline{1}11$) intensity is about 60\% of the maximal intensity, indicating that already with molecular oxygen (1.0 sccm) $m$-HfO$_{2-x}$ forms, but with lower crystal quality compared to films grown in oxygen radicals. Although the growth of hafnia using molecular oxygen is possible, lower growth pressures (lower oxygen partial pressures during growth) while maintaining comparable oxidation conditions can be realized only by using radicals instead of oxygen molecules. Furthermore, an oxygen radical source allows for a much more precise control of the oxidation conditions.

An eye-catching observation of cathodoluminescence (CL) in an oxygen deficient HfO$_{2-x}$ thin film on $r$-cut sapphire is shown in Fig.~\ref{Fig:Nice}. Irradiation with 30 keV electrons from RHEED evokes CL in HfO$_{2-x}$ due to the presence of oxygen vacancies. The CL intensity under RHEED electron beam irradiation scaled with the oxidation conditions during growth, from no CL for presumably stoichiometric films to a prominent lavender shine for highly reduced films. This increase in CL with increasing oxygen vacancy concentration is in good agreement with similar studies for SrTiO$_{3-x}$.\cite{Kalabukhov:07}


\begin{figure}[t]
\centering{%
\includegraphics[width=0.9\columnwidth,bbllx=0,bblly=0,bburx=570,bbury=667,clip=]{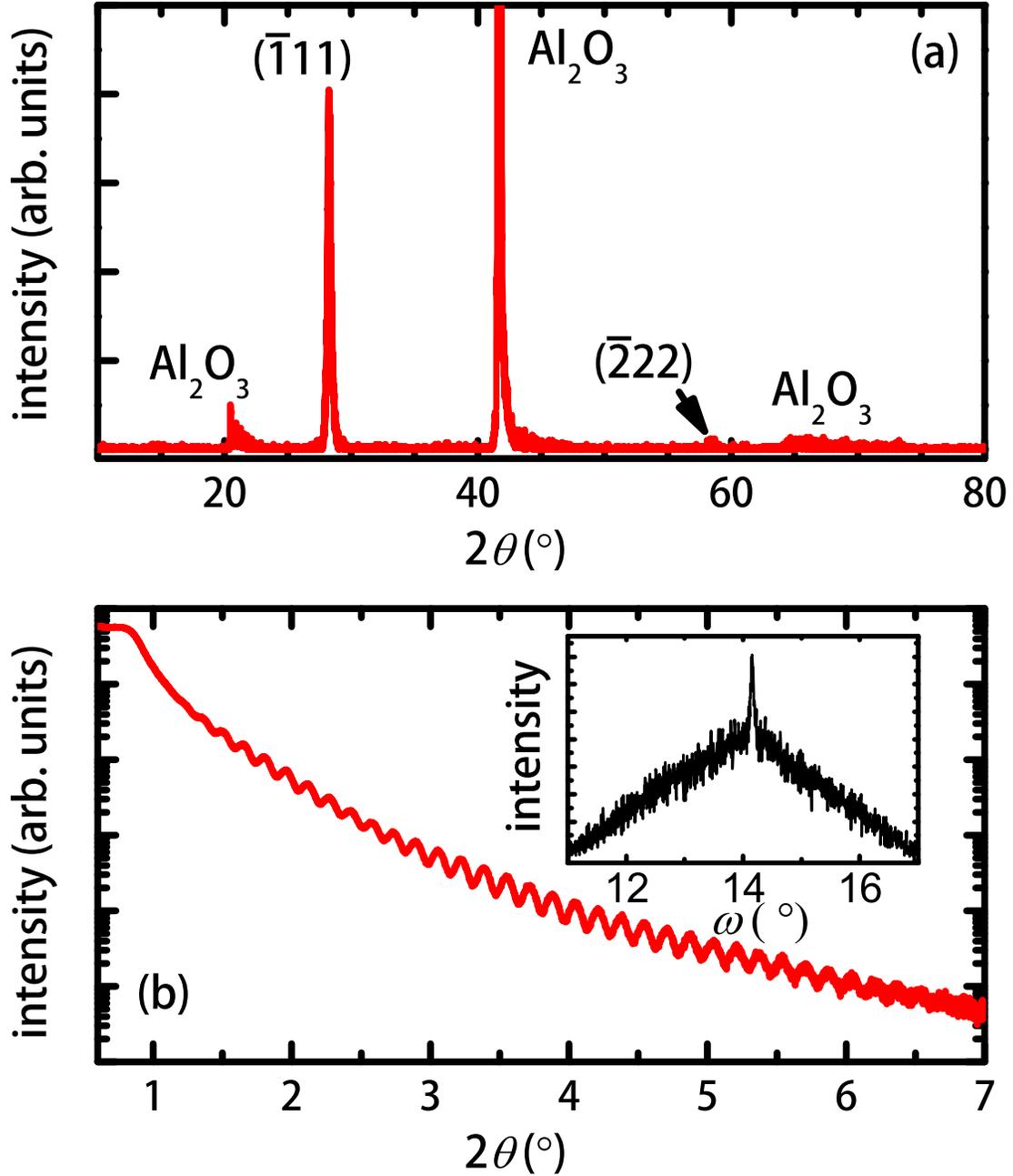}}
\caption{(Color online) (a) $2\theta-\theta$-scan for a 50 nm thick hafnium oxide thin film on $c$-cut sapphire grown under optimized conditions. (b) Corresponding XRR pattern. Inset: Rocking curve of the ($\overline{1}11$) peak.}\label{Fig:2theta}
\end{figure}

\subsubsection{Optimal growth conditions}

\begin{figure}[t]
\centering{%
\includegraphics[width=0.9\columnwidth,bbllx=0,bblly=0,bburx=508,bbury=314,clip=]{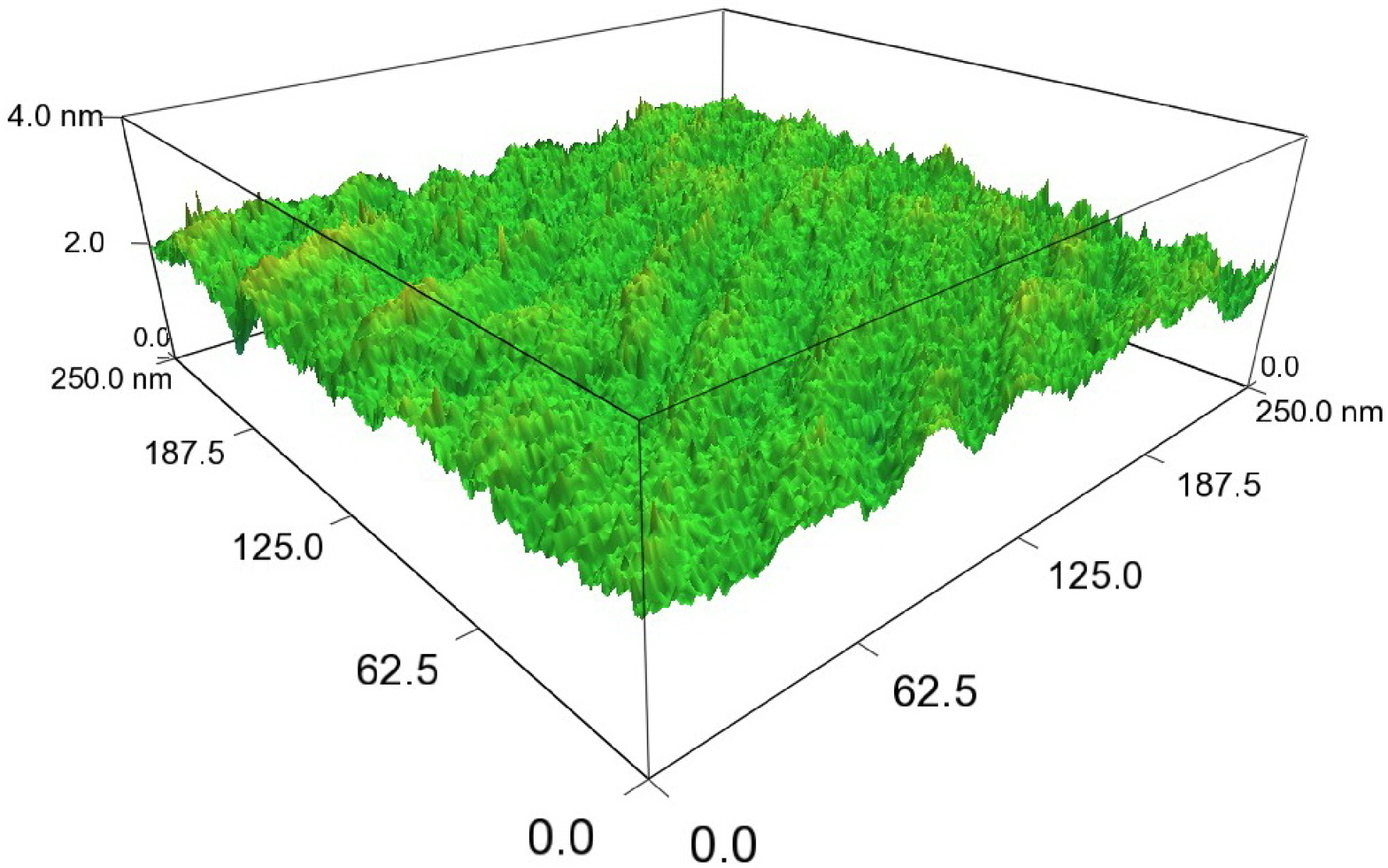}}
\caption{(Color online) AFM picture of a 50 nm thick hafnium oxide film grown under optimized conditions with a surface roughness RMS of 0.22 nm in a measured area of 1 $\mu$m$^2$.}\label{Fig:AFM}
\end{figure}

We show a $2\theta-\theta$-scan for a nominally 50 nm thick HfO$_2$ thin film on $c$-cut sapphire grown under optimized conditions with respect to substrate temperature (700 $^\circ$C), oxygen flow rate (1.0 sccm, corresponding to slight oxygen deficiency) and rf-power (200 W) in Fig.~\ref{Fig:2theta}(a). The Hf rate was 0.7 {\AA}/s. Under these conditions, a single ($\overline{1}11$) oriented monoclinic HfO$_{2-x}$ thin film is obtained. The undamped oscillations up to angels of more than 6$^\circ$ in $2\theta$  shown in the X-ray reflectivity (XRR) pattern in Fig.~\ref{Fig:2theta}(b) result from low surface and interface roughness. From the oscillation period a film thickness of 52.4 nm is extracted which is in good agreement with the calculated nominal film thickness of 50 nm. This result is further confirmed by AFM measurements giving a surface roughness less than 0.22 nm in RMS (measurement area 1 $\mu$m$^2$) as shown in Fig.~\ref{Fig:AFM}.
Similarly low surface roughness usually cannot be obtained without post deposition treatments in the case of atomic layer deposition (ALD) or metal organic MBE (MOMBE). Even thinner films than 50 nm are reported to exhibit 1.0 nm in RMS.\cite{Moon:05,Modreanu:05,Sammelselg:07} However, in the case of RMBE-grown hafnia, post deposition annealing processes are evidently not as important as for films grown using ALD or MOMBE.

\begin{figure}[b]
\centering{%
\includegraphics[width=1.0\columnwidth,bbllx=0,bblly=0,bburx=440,bbury=106,clip=]{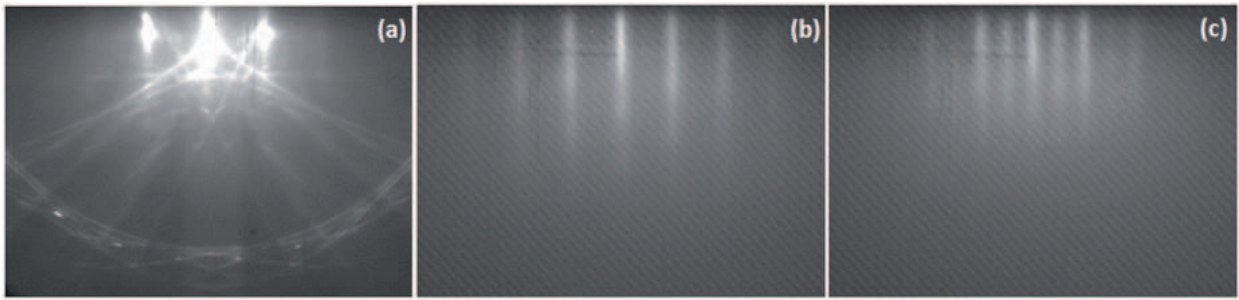}}
\caption{(Color online) RHEED images of HfO$_2$ on $c$-cut sapphire during growth; (a) blank $c$-cut substrate before growth, (b) after 3 nm, (c) after 6 nm film thickness.}\label{Fig:RHEED}
\end{figure}

We have monitored the thin film surface by RHEED which gives a rich and detailed real-time feedback of the ongoing crystal growth. In Fig.~\ref{Fig:RHEED} we show RHEED patterns obtained during the growth of optimized thin films. Clear streaks are observed up to 6 nm in film thickness, indicating initial epitaxial film growth with a smooth surface. With increasing film thickness additional streaks appear between the primary streaks corresponding to a doubling of the lattice constant. Later in film growth streaks start to vanish and spots do appear, indicating increasing film roughness. Only for very thick films ($>$ 100 nm) rings start to appear, indicating a relaxed polycrystalline film with increasing film thickness. At increased growth temperatures around 820 $^\circ$C, RHEED growth oscillations could be observed.
The doubling of the surface unit cell after growth of a few unit cells of HfO$_{2-x}$ is a novel observation. It indicates a surface reconstruction which is prevented during the growth of the first layers, but becomes stable after a certain critical film thickness of a few nm. After formation of the surface reconstruction, the further growth is hampered, and a transition to an island growth mode occurs leading to a typical transmission electron diffraction pattern.
The rocking curve of the ($\overline{1}11$) hafnia peak as shown in the inset of Fig.~\ref{Fig:2theta}(b) indicates two contributions. One sharp peak associated with the epitaxial layer growing directly at the substrate film interface with a typical FWHM of a few tenth of degree. The broad peak is due to the relaxed part of the film which forms after the epitaxy of the initial layers is lost. After relaxation the FWHM amounts to several degrees.
In summary, the resulting thin films are highly textured in ($\overline{1}11$) orientation and relaxed, as can be seen from X-ray diffraction. However, due to the lack of in-plane registry with the substrate, the films are not epitaxial. As shown by high-resolution transmission electron microscopy (HRTEM) the films consist of highly textured nanocrystalline regions.\cite{Hildebrandt:11}


\begin{figure}[t]
\centering{%
\includegraphics[width=1.0\columnwidth,bbllx=0,bblly=0,bburx=576,bbury=412,clip=]{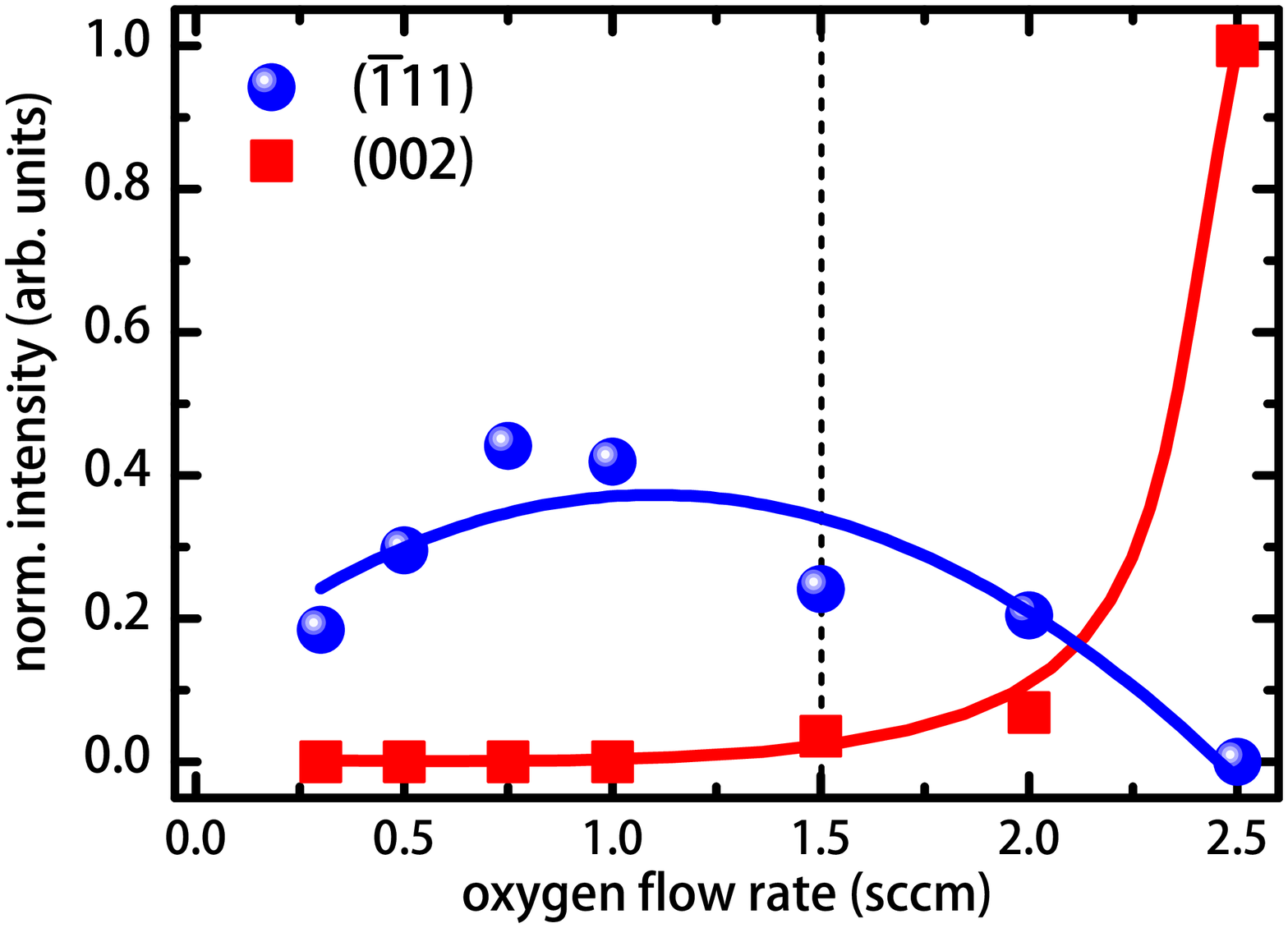}}
\caption{(Color online) Normalized intensities of ($\overline{1}11$) and ($002$) reflections of 200 nm thick hafnia thin films as a function of oxygen flow rate (Hf-rate 0.7 {\AA}/s, $T_{\text{S}} = 700 ^\circ$C, 200 W rf-power for 0 - 2.0 sccm, 300 W rf-power for 2.5 sccm). Solid lines are guides to the eye. The vertical dashed line indicates approximately the oxygen flow rate corresponding to stoichiometric composition.}\label{Fig:OxFlow}
\end{figure}

\subsection{Film orientation as a function of oxygen content}

Now, we focus on the influence of oxygen stoichiometry on thin film orientation. For this purpose, thin films of 200 nm thickness with oxygen flow rates ranging from 0.25 to 2.5 sccm have been grown and characterized with $2\theta$-$\theta$ out-of-plane measurements. Other deposition parameters were 0.7 {\AA}/s Hf-rate, 700 $^\circ$C substrate temperature, and 200 to 300 W of rf-power. The film thickness was increased in order to obtain more volume fraction for X-ray diffraction. Fig.~\ref{Fig:OxFlow} shows the variation of the normalized intensities for the ($\overline{1}11$) and ($002$) reflections of $m$-HfO$_2$ on $c$-cut sapphire substrates. The intensity of the ($\overline{1}11$) reflection shows a maximum in the regime of 0.7 sccm oxygen flow rate, whereas the vanishing intensity of the ($002$) reflection shows no significant variation up to 1.0 sccm oxygen flow rate. At higher flow rates the ($002$) intensity slowly picks up, and shows a steep increase for 2.5 sccm and 300 W rf-power. Note that for the sample grown with 2.5 sccm oxygen flow rate the rf-power applied to the radical source has been increased from 200 to 300 W to investigate very high oxidation conditions. Thus, under highest oxidation conditions the ($002$) orientation becomes dominant, while the ($\overline{1}11$) is fully suppressed.

\begin{figure}[t]
\centering{%
\includegraphics[width=1.0\columnwidth,bbllx=0,bblly=0,bburx=576,bbury=382,clip=]{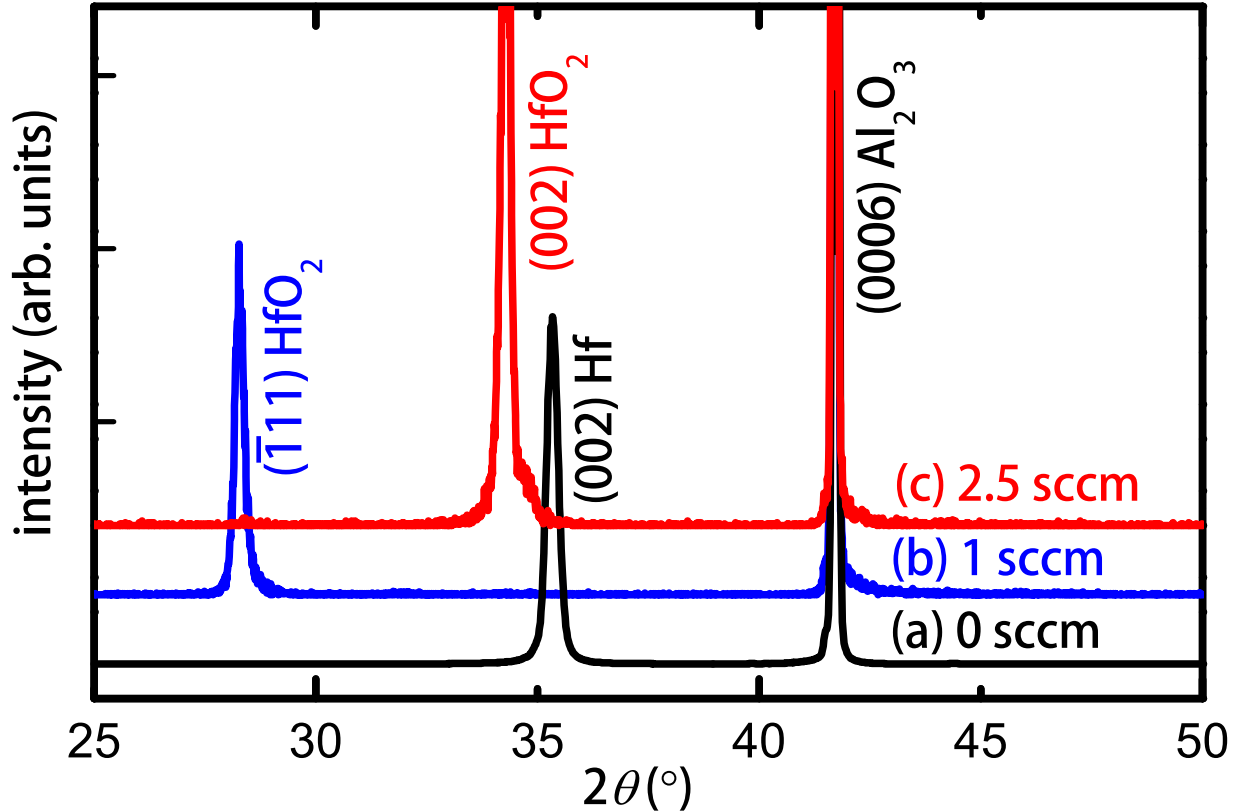}}
\caption{(Color online) Diffraction patterns for 200 nm thick hafnia thin films and 50 nm metal Hf demonstrating switching of film orientation as a function of oxidation conditions. (a) 0 sccm oxygen flow rate (metallic Hf); (b) 1.0 sccm, 200 W; (c) 2.5 sccm, 300 W rf-power. Other deposition parameters were 0.7 {\AA}/s Hf-rate, and $T_{\text{S}} = 700 ^\circ$C.}\label{Fig:Hf}
\end{figure}

As stated before, in the case of oxidation by oxygen radicals even for extreme low oxidation conditions the formation of metallic hafnium was never observed. Only when utilizing molecular oxygen instead of oxygen radicals, metallic hafnium forms for flow rates below 0.2 sccm. For zero oxygen flow rate, pure metallic hafnium stabilizes in its hexagonal phase in ($002$) orientation as shown in Fig.~\ref{Fig:Hf}. As evident from the figure, thin film orientation can be switched as a function of oxidation conditions from ($\overline{1}11$)-oriented films obtained for 0.7 - 1.0 sccm oxygen flow rate to ($002$)-oriented films grown under highly oxidizing conditions (2.5 sccm, 300 W rf-power). In no case, any other crystalline phases of hafnia could be observed, regardless of its oxygen content. Reports on the stabilization of thermodynamically unstable phases like cubic or tetragonal HfO$_2$ under ambient conditions utilising high-energy ion beam assisted deposition could not be reproduced.\cite{Mori:03,Miyake:01} In addition to $c$-cut sapphire substrates, also $r$-cut sapphire substrates were used. However, on these substrates slightly rougher, more polycrystalline, and mostly randomly oriented films of $m$-HfO$_{2±x}$ were obtained. Because of this, further studies were limited to $c$-cut sapphire substrates.


\subsection{Band gap evolution with oxygen stoichiometry}

\begin{figure}[t]
\centering{%
\includegraphics[width=1.0\columnwidth,bbllx=0,bblly=0,bburx=455,bbury=204,clip=]{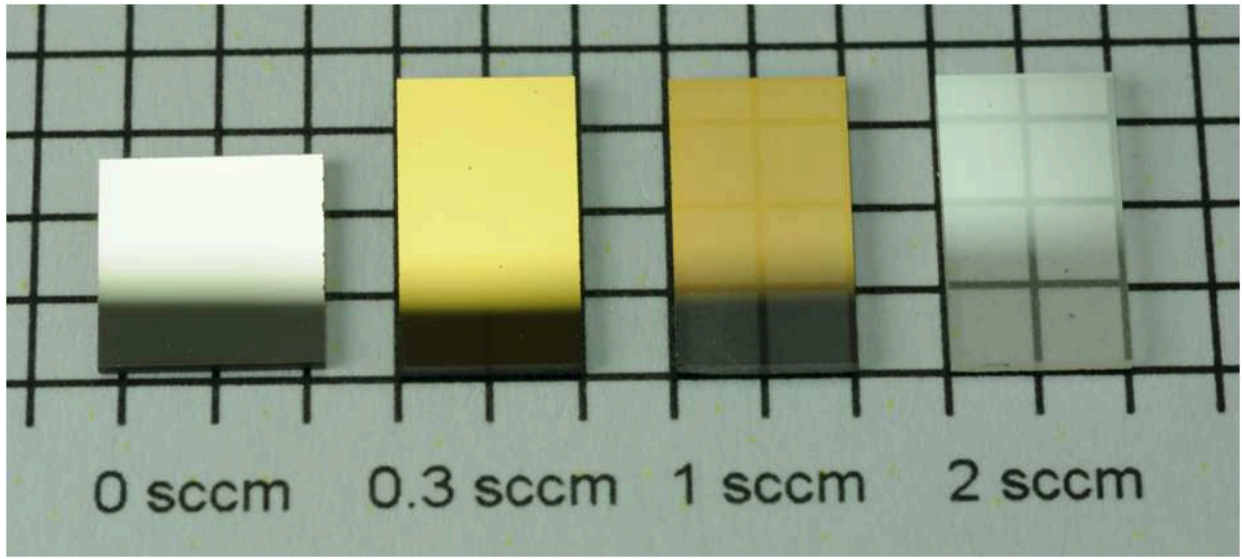}}
\caption{(Color online) Photograph of three HfO$_{2-x}$ thin films: 0.3 sccm oxygen flow rate for highly reduced growth conditions; 1 sccm for reduced, and 2 sccm for moderate oxidation conditions. As comparison also a metallic Hf film is shown (0 sccm).}\label{Fig:beauty}
\end{figure}

In this section we focus on the evolution of the optical band gap as a function of oxygen content in HfO$_{2±x}$ thin films. We start with a palpable observation in the visible
wavelength range. When having a look on the grown 200 nm films, one notices a darkening and change of color of the films as a function of decreased oxidation conditions during growth.
In Fig.~\ref{Fig:beauty} three hafnium oxide thin films grown under different oxidation conditions are shown. A flow rate of 2.0 sccm leads to transparent (close to stoichiometric) films, 1.0 sccm leads to slightly darker and less transparent films, and 0.3 sccm leads to dark, non-transparent thin films with a beautiful golden shine. As a reference, a 50 nm thick, purely metallic hafnium film is added to Fig.~\ref{Fig:beauty} (corresponding to 0 sccm). Film darkening in hafnia as a function of oxidation conditions during growth, to our knowledge, has only been observed before by Hadacek {\em et al.}, where a slight grey appearance for films grown under UHV conditions with PLD has been reported.\cite{Hadacek:07} In comparison, for oxygen deficient SrTiO$_{3-\delta}$, a change in color as a function of oxygen deficiency has been reported, where a glistening oxidized gem transmutes into a dull blue, conductive crystal.\cite{Mannhart:04} We will later discuss the color change in HfO$_{2-x}$ as a consequence of the formation of mid-gap defect bands due to oxygen vacancies allowing absorption at energies above 2 eV (620 nm).


\begin{figure}[t]
\centering{%
\includegraphics[width=1.0\columnwidth,bbllx=0,bblly=0,bburx=576,bbury=664,clip=]{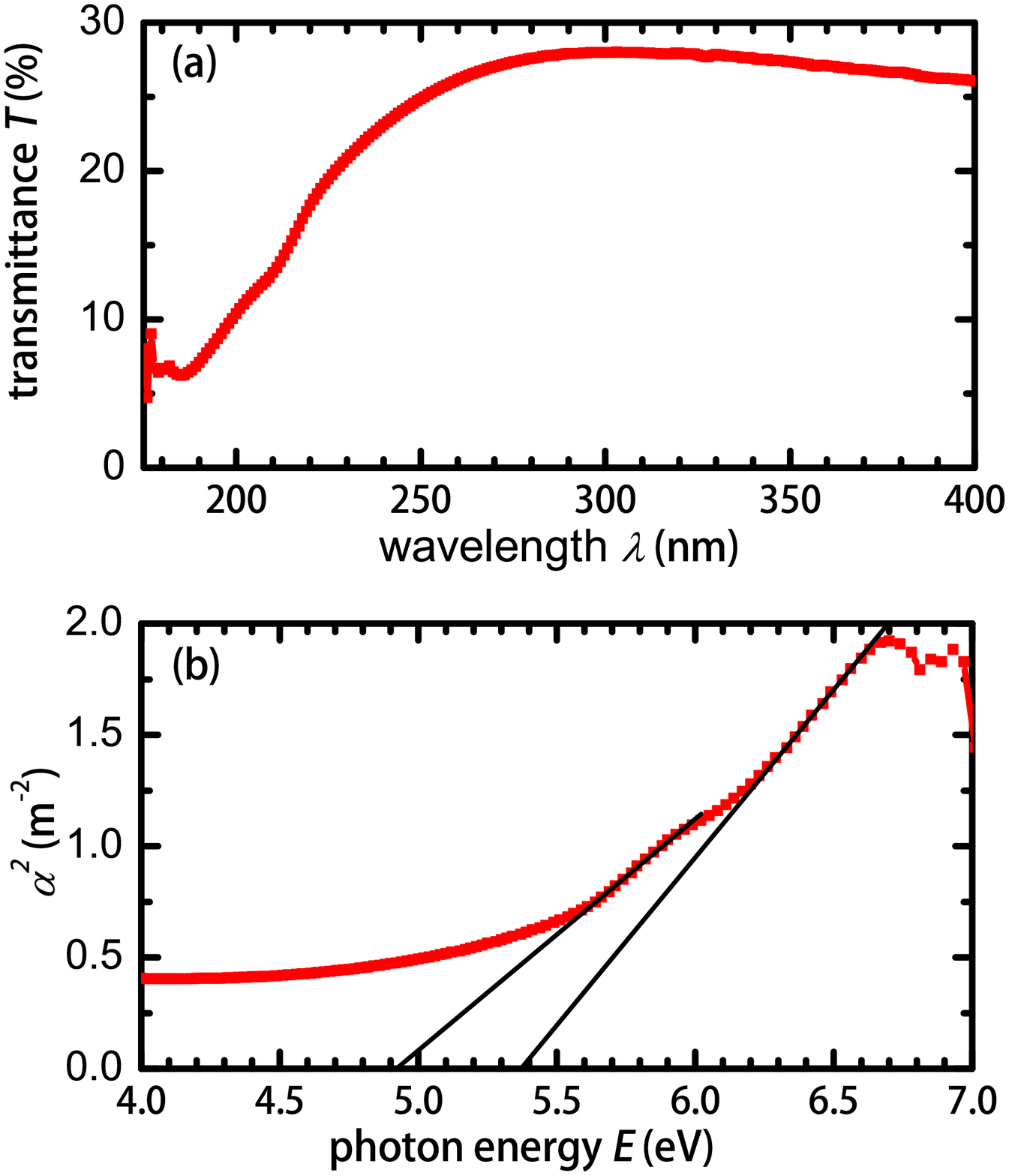}}
\caption{(Color online) (a) Transmission vs wavelength spectrum for a 30 nm thick HfO$_{2-x}$ thin film grown under reducing conditions. (b) Corresponding squared absorption coefficient $\alpha^2$ vs. photon energy.}\label{Fig:trans}
\end{figure}

Stoichiometric HfO$_2$ is an insulator and a wide band gap material with a band gap of around 5.7 eV.\cite{Robertson:02} Lots of studies regarding the band gap of hafnium oxide have been conducted. A recent overview of the applied techniques is presented by Cheynet {\em et al}.\cite{Cheynet:07} However, the focus of most studies is the investigation of absorption features near the absorption edge of the optical band gap and the comparison between single crystalline, polycrystalline, and amorphous thin films.\cite{Aarik:04,Martinez:07} For this study, absorption measurements using a photospectrometer in transmission geometry have been conducted. In order to reduce absorption due to molecular oxygen and water vapor for wave lengths below 200 nm, the sample chamber has been flushed continuously with gaseous nitrogen. For the determination of the reflected portion $R$ and the absorbed portion $A_{\text{sub}}$ resulting from the sapphire substrates, background measurements with bare substrates have been performed. Additionally, films with thicknesses of 30 nm were used instead of 200 nm, because, first, the transmitted intensity for highly reduced samples is too low to allow reasonable measurements in the case of 200 nm thick films, and, second, unwanted interferences due to vicinity of the radiation wavelength and film thickness can be reduced to a minimum. Figure~\ref{Fig:trans}(a) shows exemplarily a transmission spectra obtained for a reduced HfO$_{2-x}$ thin film in the range of 175 to 400 nm. The transmittance for wavelengths of 300 nm is reduced from values of around 100\% for stoichiometric HfO$_2$ to values of 25\%. The optical band gap determination was carried out by extrapolating the linear parts of the $\alpha^2$ vs. radiation energy function considering direct band gap, parabolic valence, and conduction bands.\cite{Rincon:84,Aqili:02} Fig.\ref{Fig:trans}(b) exemplarily shows the absorption coefficient, $\alpha$, dependence on irradiation energy for a 30 nm thick HfO$_{2-x}$ thin film grown under reducing conditions. Two separate absorption features can be identified, extrapolating to 5.38 and 4.94 eV. The presence of two features near the absorption edge has been reported earlier for sputtered hafnium oxide thin films studied as a function of the oxygen to argon ratio.\cite{Hoppe:07} However, these features did not show a dependence on oxygen to argon ratio, indicating that the investigated films were not of sufficient oxygen deficient nature to give raise to a significant change of the band gap, $E_{\text{g}}$. In this study, a remarkable narrowing of the band gap of oxygen deficient HfO$_{2-x}$ of more than 1 eV from 5.7 eV in the stoichiometric case to values as low as 4.5 eV was observed. A change in optical band gap as a function of oxidation conditions in a limited range of 0.2 eV has been found for sputtered hafnium oxide,\cite{Pereira:05,Park:08} post deposition annealed PLD-grown hafnium oxide,\cite{Zheng:06} and titanium dioxide.\cite{Tomaszewski:02} As discussed later, the band gap reduction is probably due to a hybridization of in-gap defect states with the conduction band.


\begin{figure}[t]
\centering{%
\includegraphics[width=1.0\columnwidth,bbllx=0,bblly=0,bburx=576,bbury=416,clip=]{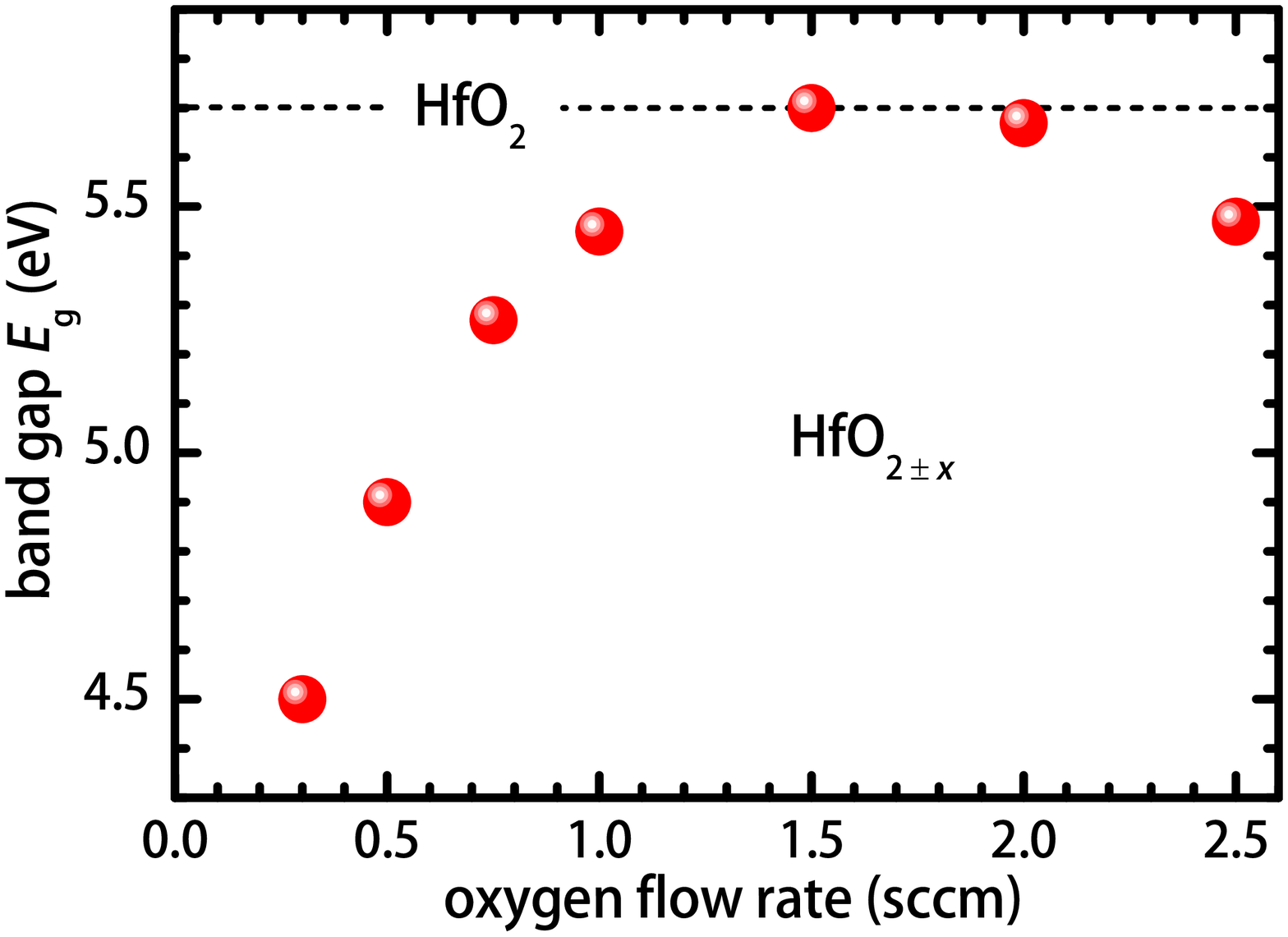}}
\caption{(Color online) Band gap $E_{\text{g}}$ vs. oxygen flow rate of HfO$_{2\pm x}$ thin films. The horizontal line indicates 5.7 eV for stoichiometric HfO$_2$.}\label{Fig:gap}
\end{figure}

Figure~\ref{Fig:gap} shows $E_{\text{g}}$ as a function of oxidation conditions represented by the oxygen flow rate. A parabolic dependence with a maximum around 5.7 eV for stoichiometric HfO$_2$ is found. The observed decrease in band gap results either from oxygen vacancies in HfO$_{2-x}$ or from hafnium vacancies/oxygen interstitials in HfO$_{2+x}$. These observations point to the evolution of in gap defect bands which reduce the total band gap by hybridization with the conduction or valence band. This scenario has been simulated for isolated defect states by density functional theory.\cite{Xiong:05,Pemmaraju:05} The presence of defect states implies charge carriers which can be localized. The transition to extended defect states respective defect bands implies electric conductivity. The introduction of oxygen vacancies usually is associated with electron doping, as for example observed for SrTiO$_3$.\cite{Kalabukhov:07}

\subsection{Electrical properties of HfO$_{2-x}$}

The introduction of oxygen vacancies in HfO$_{2-x}$ creates charge carriers in the thin films. Stoichiometric HfO$_2$ is a well-known, transparent insulator with a resistivity in the range of 10$^{18}$ $\mu\Omega$cm. In literature, the variation of resistivity vs. oxidation conditions during growth has been studied in the high-resistivity regime above 10$^{15}$ $\mu\Omega$cm showing a decrease in resistivity for non-stoichiometric thin films.\cite{Pereira:05} Bharathi {\em et al}. observed resistivities in the range of 10$^9$ $\mu\Omega$cm as a function of oxidation conditions for reactive sputtered films on quartz and stripped Si (100) substrates grown at temperatures up to 500 $^\circ$C.\cite{Bharathi:10} Their study describes a variation in conductivity of about 50\% as a function of crystallite size. According to the presented absorption data and compared to our study, those samples are only slightly oxygen deficient. A formation of hafnium based silicates due to diffusion could be a possibility to explain the low resistivity. This would explain the missing ($001$)-reflection for samples grown at elevated temperatures. In our case, for films grown under lower oxidation conditions with decreased transparency and narrowed optical band gap, resistivities in the range of 10$^7$ down to a couple of hundred $\mu\Omega$cm were measured.

\begin{figure}[t]
\centering{%
\includegraphics[width=1.0\columnwidth,bbllx=0,bblly=0,bburx=573,bbury=410,clip=]{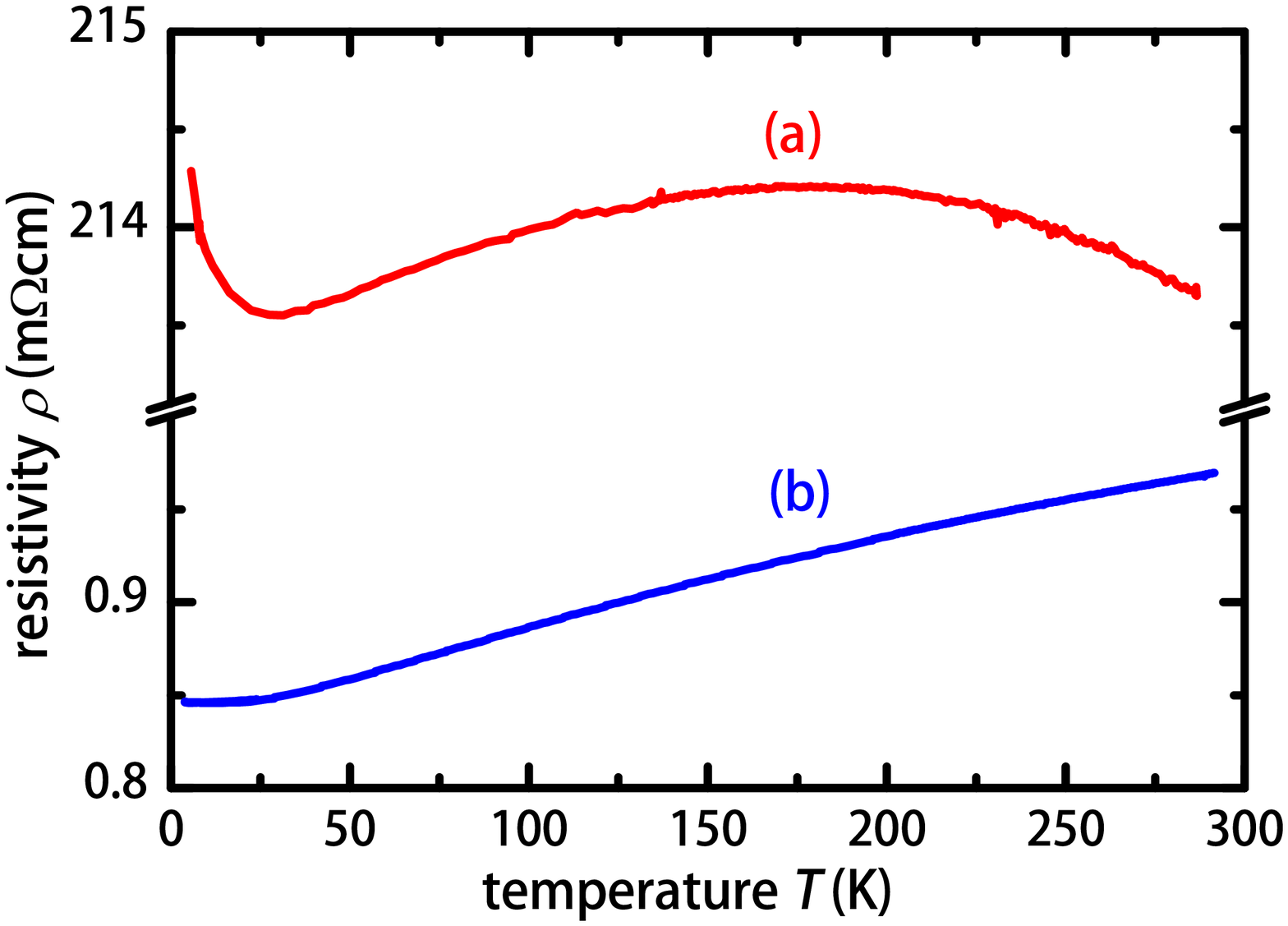}}
\caption{(Color online) Resistivity vs. temperature for 50 nm thick HfO$_{2-x}$ thin films grown (a) under slightly reducing ($0.9$ sccm oxygen flow), and (b) under highly reducing condition ($0.6$ sccm oxygen flow)).}\label{Fig:rhoT}
\end{figure}

\begin{figure}[t]
\centering{%
\includegraphics[width=1.0\columnwidth,bbllx=0,bblly=0,bburx=576,bbury=417,clip=]{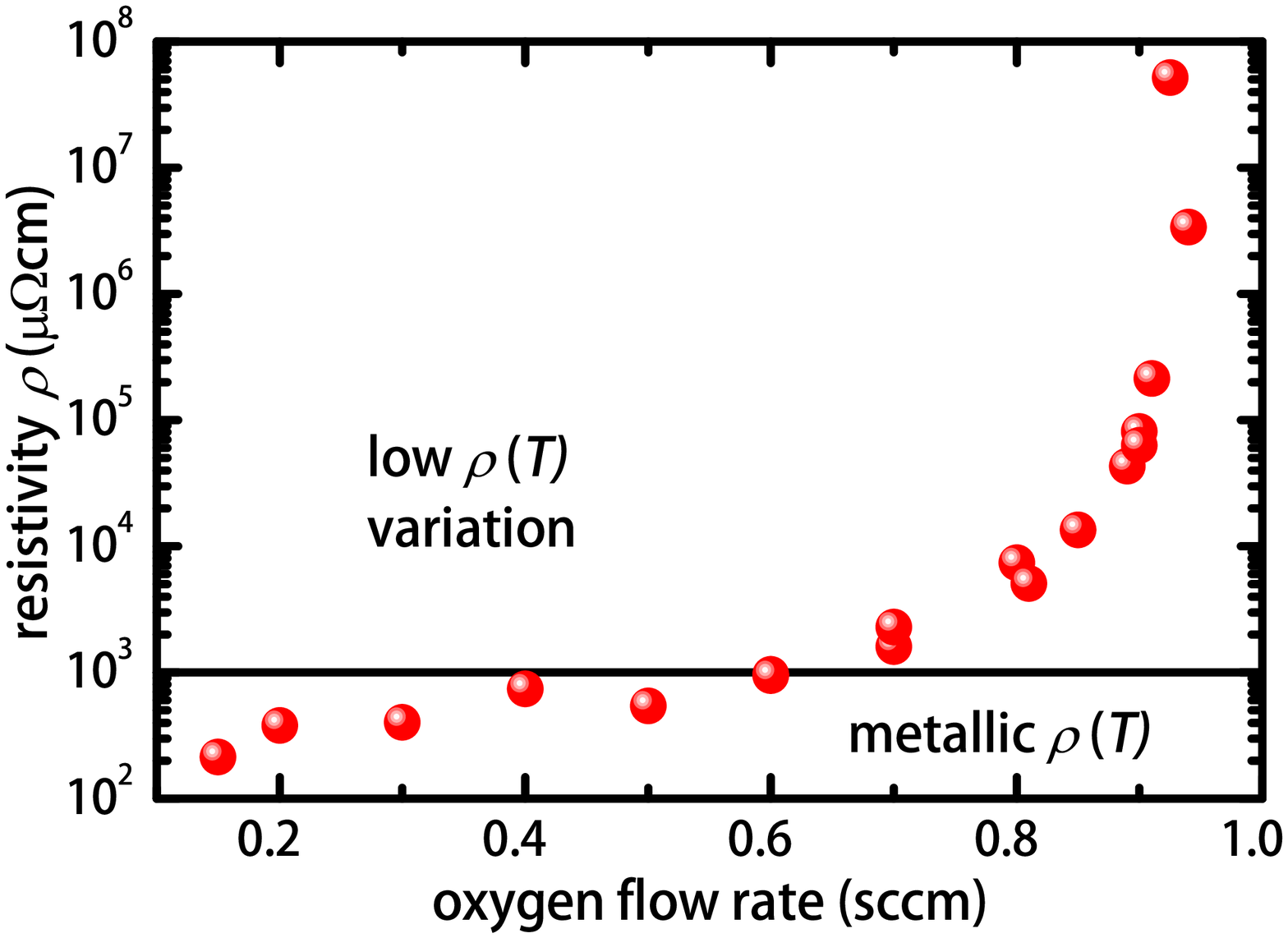}}
\caption{(Color online) Room-temperature resistivity as a function of oxidation conditions. Films grown below 0.6 sccm oxygen flow rate show metallic $\rho(T)$, whereas above this threshold no major change in $\rho$ as a function of $T$ is observed.}\label{Fig:rho}
\end{figure}

Figure~\ref{Fig:rhoT} shows the thin film resistivity as a function of temperature, $\rho(T)$.
For slightly reduced samples, we observe a resistivity in the range of $200,000$ $\mu\Omega$cm
which is almost temperature independent. At low temperatures, an upturn of the resistivity indicates a freezing of the charge carriers. Highly reduced samples show a clear metallic $\rho(T)$ behavior. The resistivity starts at about 1 m$\Omega$cm at room temperature and drops by around 12\% to a residual resistivity of 840 $\mu\Omega$cm at 5 K. The resistivity of pure, bulk metallic hafnium is in the range of 32 $\mu\Omega$cm, well below the observed values for HfO$_{2-x}$. When having a closer look on the variation of room temperature resistivity as a function of oxidation conditions, one can observe an exponential increase of resistivity as a function of oxidation conditions, and, thus, as a function of oxygen content in HfO$_{2-x}$ thin films. This trend is shown in Fig.~\ref{Fig:rho}.

One scenario which could give an explanation for such a tremendous decrease in resistivity in oxygen deficient HfO$_{2-x}$ thin films of more than 14 orders of magnitude is the possible presence of metallic hafnium distributed in the film forming conductive current paths. A volume fraction of more than 20\% of metal hafnium would be needed, assuming a statistical distribution in an HfO$_2$ matrix according to percolation theory.\cite{Shante:71} In order to investigate this further, X-ray measurements were carried out to trace crystalline metallic hafnium. In all cases, monoclinic hafnium oxide was identifiable. No impurity or metal Hf phases could be identified. However, there is still the possibility for a distribution of amorphous metallic hafnium in a stoichiometric HfO$_2$ matrix which would not be observable utilizing X-ray diffraction. To rule out this scenario, HRTEM has been applied. With this method, one could clearly identify HfO$_{2-x}$ grains with corresponding atomic spacings, but no traces of amorphous phases. Additionally, both, X-ray diffraction and HRTEM, show that even for extreme low oxidation conditions still the monoclinic structure of hafnia is maintained without decomposition or phase change. A more detailed description of the HRTEM study can be found elsewhere.\cite{Hildebrandt:11} All films were stable in air and were re-measured after 12 months, showing no change of physical properties.


As evident from the resistivity measurements, the introduction of oxygen vacancies leads to the formation of charge carriers in HfO$_{2-x}$ thin films, thus leading to conductivities exceeding values known for other semiconductors.\cite{Dietl:00} To clarify the nature of charge carriers in the oxygen deficient films, Hall measurements in van-der-Pauw-geometry at 300 and 5 K have been carried out utilizing a MPMS dewar with a field capability of $\pm$ 7 Tesla. The obtained Hall voltage as a function of the applied magnetic field follows a linear behavior. No anomalous Hall effect could be observed. A positive Hall coefficient of $R_{\text{H}} = 8.7\times10^{-4}$ cm$^3$/C indicates holes as majority charge carriers with concentrations of $6\times10^{21}$ cm$^{-3}$ and mobilities of 2 cm$^2$/(Vs) for a 50 nm thick HfO$_{2-x}$ film grown under 0.25 sccm oxygen flow rate. No change of the Hall coefficient, no significant change in charge carrier concentration, and charge carrier mobility measured at 300 and 5 K could be observed. The observed $p$-type conductivity is in agreement with earlier findings, claiming oxygen vacancies being responsible for electrical conductivity in HfO$_{2-x}$.\cite{Hadacek:07} In contrast, Ko {\em et al}. emphasize the importance of charged point defects (oxygen interstitials or hafnium vacancies) responsible for high temperature conductivity, rather than oxygen vacancies.\cite{Ko:10} Taking the charge carrier concentration of $6\times10^{21}$/cm$^3$ and the assumption, that each oxygen vacancy traps one electron, one can roughly estimate the stoichiometry of this particular film to be $x = 0.2$. In literature, similarly stoichiometry  values have been reported, however, in all these cases neither high conductivity nor change in optical appearance was reported.\cite{Miyake:01,Capone:98,Hong:06} It would be favorable to determine absolute values of oxygen stoichiometry, avoiding indirect methods as our estimation using the charge carrier concentration in highly deficient HfO$_{2-x}$. Various approaches have been made to determine the oxygen stoichiometry, in many cases X-ray photoelectron spectroscopy has been applied to monitor the O $1s$ and Hf $4f$ band intensities and positions.\cite{Cho:07,Liu:11} These XPS studies can be carried out {\em in situ} or {\em ex situ}, in the latter case with or without sputtering for cleaning and/or depth profiling. To determine the stoichiometry of our RMBE grown thin films, sputtered XPS has been applied, but considered not to be accurate enough due to significant preferential sputtering of oxygen. This leads to a measured decrease in oxygen content with increasing sputtering time, as evident from depth profiles. This effect becomes more and more prominent, the higher the film deficiency is. The determination of $x$ via {\em ex situ} XPS without sputtering is difficult because of the formation of a surface layer of fully oxidized HfO$_2$. This surface oxidation occurs when the samples are taken out of the growth chamber. We have observed a shift of the O $1s$ signal as a function of increasing oxidation conditions to higher binding energies, which is in good agreement with prior findings in literature.\cite{Moon:05} Miyake {\em et al}. report the determination of $x$ in HfO$_{2-x}$ via XPS resulting in HfO$_{1.535}$,\cite{Miyake:01} Capone {\em et al}. report even lower values of $x = 1.454$.\cite{Capone:98} These values do not match well to our results, as from the amount of charge carriers our estimated deficiency is in the range of $x = 0.2$ which are more conservatively estimated avoiding unreliable results of oxygen XPS.

\subsection{Magnetic properties of HfO$_{2-x}$}

\begin{figure}[t]
\centering{%
\includegraphics[width=1.0\columnwidth,bbllx=0,bblly=0,bburx=576,bbury=390,clip=]{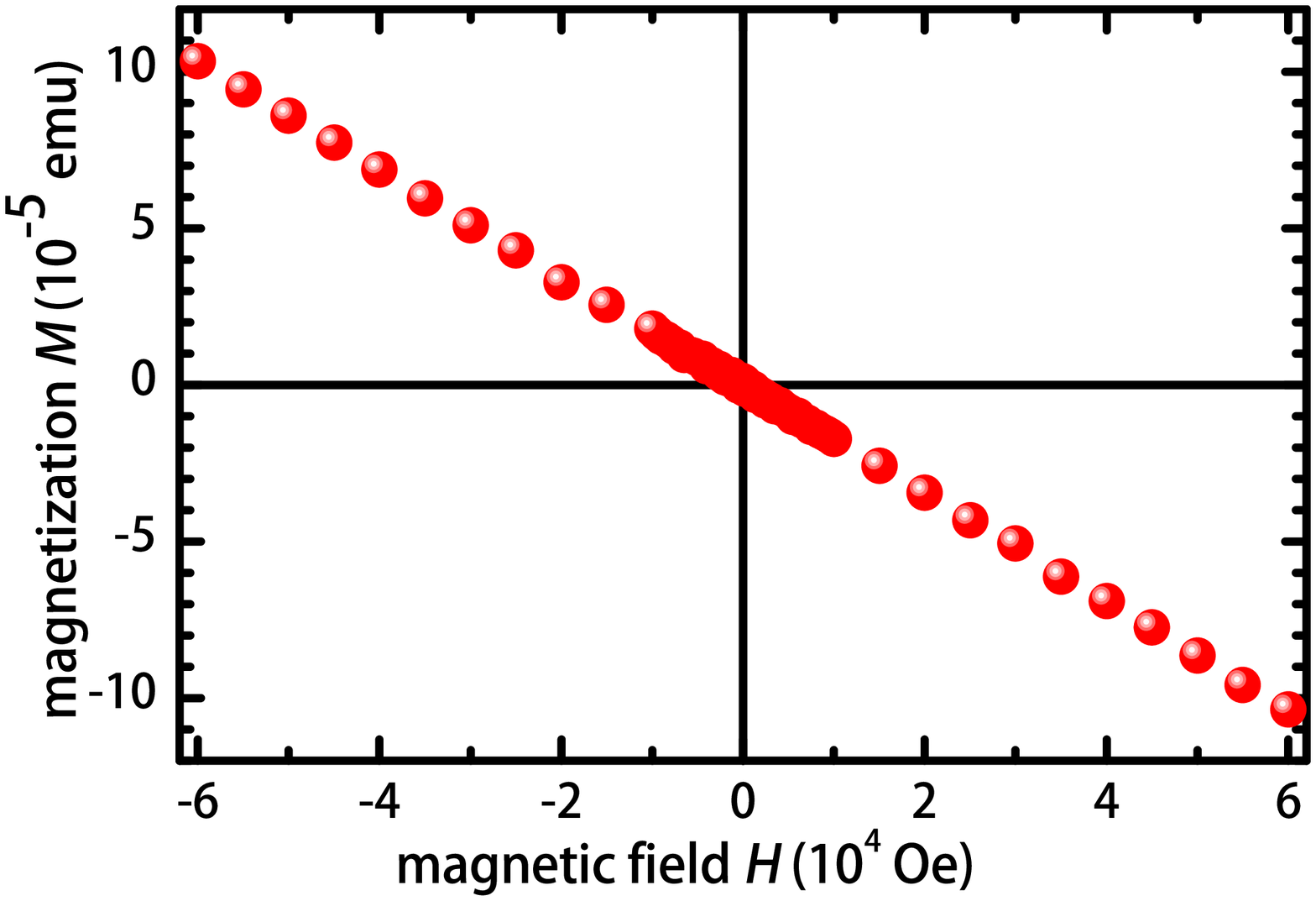}}
\caption{(Color online) Magnetization data obtained at 300 K for a 200 nm thick HfO$_{2-x}$ thin film grown under 0.3 sccm oxygen flow rate on $c$-cut sapphire. The diamagnetic substrate contribution is not subtracted. All investigated films show diamagnetic behavior, regardless of the oxidation conditions.}\label{Fig:mag}
\end{figure}

Based on the results described in Secs. IIIA-IIID, oxygen vacancies are introduced into hafnium oxide thin films grown by RMBE in a controlled and reproducible manner. Consequently, this allows the investigation of possible $d^0$-ferromagnetism in undoped transition metal oxides, which is attributed to the presence of oxygen vacancies\cite{Venkatesan:04,Coey:05,Bouzerar:06} Recently, supplementary to the claim of $d^0$-ferromagnetism, room temperature ferromagnetism with dopants below the percolation limit, such as, Ni,\cite{Sharma:10} Co,\cite{Soo:07}, and Fe,\cite{Hong:06} has caused significant interest in the scientific community. In this study, the investigation in terms of possible ferromagnetism will concentrate on dopant-free, oxygen deficient hafnia. Prior to deposition, the composition of the used hafnium source and after deposition the resulting thin films have been investigated by secondary ion mass spectrometry (SIMS) in order to check for any above mentioned magnetic impurities. In none of the measurements traces of Fe, Ni, or Co could be identified. After deposition, magnetisation data of a series of oxygen deficient HfO$_{2-x}$ thin films has been obtained with a MPMS at various magnetic fields and temperatures. Samples brought in contact with metallic tools can lead to a ferromagnetic feature as reported by Abraham {\em et al}.\cite{Abraham:05} Extreme care in sample handling was applied to avoid unwanted impurities of any kind. Fig.~\ref{Fig:mag} shows exemplarily the magnetization curve for a 200 nm thick hafnium oxide thin film grown under highly reducing conditions (0.3 sccm oxygen flow rate) obtained at 300 K. Note that the diamagnetic contribution from the sapphire substrate has not been subtracted. Clearly, the linear diamagnetic behavior known for diamagnetic and stoichiometric HfO$_2$ could be observed, with no traces of ferromagnetic features. This is in good agreement with results found in literature claiming the absence of $d^0$-ferromagnetism for undoped transition metal oxides.\cite{Hadacek:07} For this study numerous films have been deposited and regularly measured, all results fail to give any evidence of $d^0$-ferromagnetism in undoped hafnia. This finding stands for films grown under high and low oxidation conditions with oxygen radicals (oxygen flow rate between 0.2 and 2.5 sccm), including films showing a decreased band gap and very low resistivities. However, ferromagnetism in samples combining well controlled oxygen stoichiometry and magnetic dopants cannot be excluded by our present study.


\begin{figure}[t]
\centering{%
\includegraphics[width=0.9\columnwidth,bbllx=0,bblly=0,bburx=429,bbury=386,clip=]{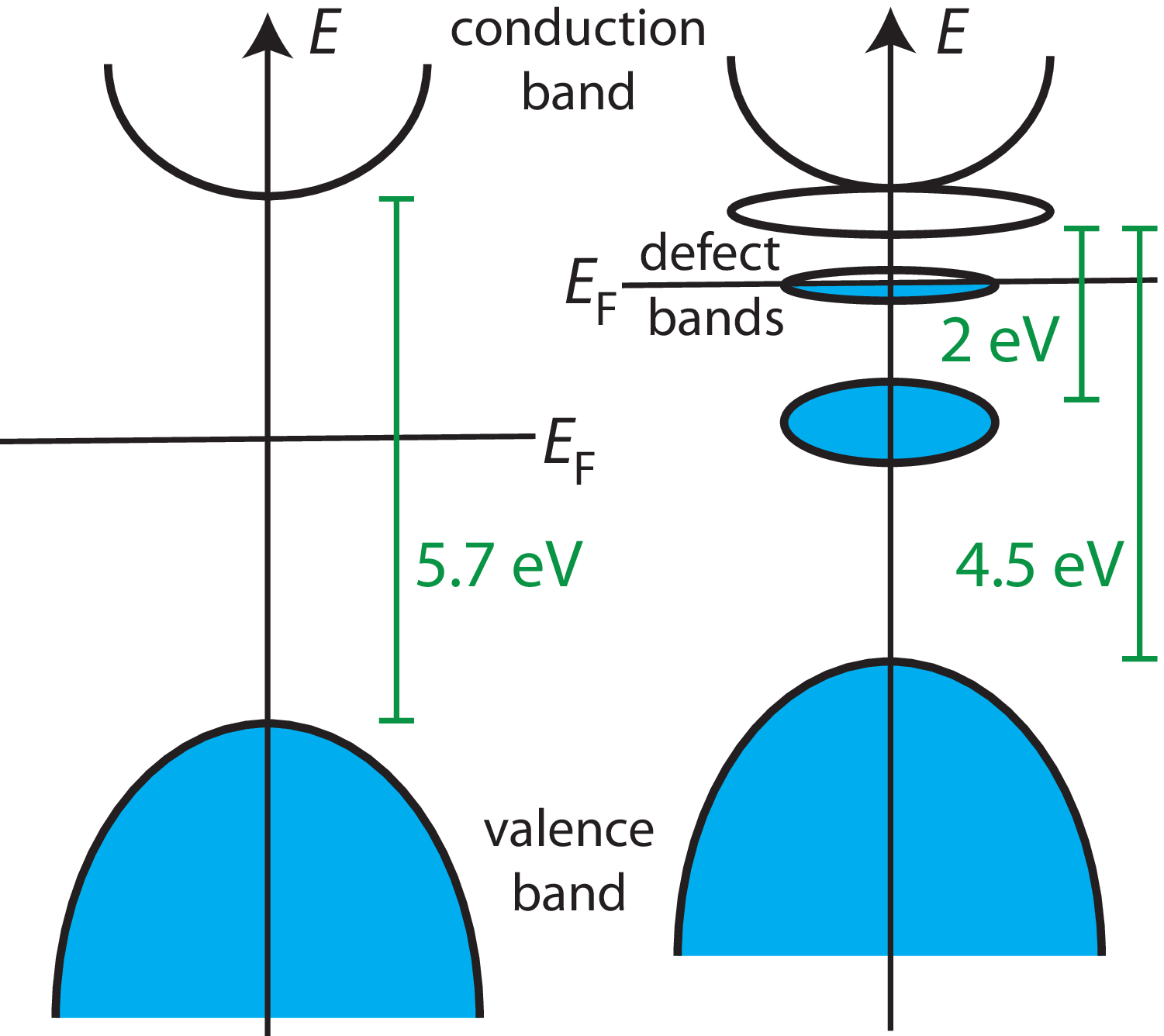}}
\caption{(Color online) Left side: Band model of stoichiometric HfO$_2$. Right side: Simplified band model of oxygen deficient HfO$_{2-x}$. }\label{Fig:model}
\end{figure}

\section{Band model of oxygen deficient HfO$_{2-x}$}

In order to understand the observed electrical conductivity, the decreasing optical band gap as a function of decreasing oxidation conditions, and the appearance of additional features in the  $\alpha^2$ vs. photon energy plots with decreasing oxidation conditions, we have developed a simple band structure model. Our model is based on the existing knowledge of individual defect states in HfO$_2$ which has not yet been extended to high defect concentrations. There are mainly three different possible vacancies in hafnia, three-fold and four-fold coordinated oxygen vacancies, and a seven-fold coordinated hafnium vacancy. Oxygen vacancies in hafnia can be double or single positively charged ($V_{\text{O}}^{++}$, $V_{\text{O}}^{+}$), neutral ($V_{\text{O}}^{0}$), and single or double negatively charged ($V_{\text{O}}^{-}$, $V_{\text{O}}^{--}$), dependent on the number of trapped electrons. The corresponding energy levels have been calculated using the screened exact exchange and weighted density approximation methods within the local density formalism.\cite{Xiong:05} These methods have the advantage to give the correct band-gap of stoichiometric HfO$_2$. Similar results have been obtained by several authors.\cite{Munoz-Ramo:07a,Munoz-Ramo:07b,Broqvist:06,Pemmaraju:05} One-electron level differences only provide a poor estimate of optical transition energies, therefore, time dependent DFT (TDDFT) has been applied leading to different absorption energies.\cite{Munoz-Ramo:07b} As an example, for $V_{\text{O}}^{-}$ states the transition energy from occupied mid-gap states to the conduction band is calculated to be 3.54 eV via the one-electron level, whereas for TDDFT this energy is calculated to be 3.20 eV. In the case of low defect concentrations, the model of discrete defect states and corresponding energies is valid, whereas for very high defect concentrations as in the present study, defect states are likely to overlap and form defect bands. Calculations of the formation energies for the different mentioned oxygen vacancies as a function of Fermi energy predict that the single negatively charged defect $V_{\text{O}}^{-}$ is favored compared to the other four oxygen vacancy defects for a Fermi level placed near the conduction band.\cite{Xiong:05,Broqvist:06} From these calculations we assume that for highly oxygen deficient HfO$_{2-x}$, the $V_{\text{O}}^{-}$ is the predominant defect. The overlapping of the defect states leads to two defect bands, one fully filled mid-gap band and one partially filled band in close vicinity of the conduction band, according to the one-electron level energy calculation.

In Fig.~\ref{Fig:model} we have summarized the band structure of oxygen deficient HfO$_{2-x}$ based on our experiments. The large band gap of stoichiometric HfO$_2$ of 5.7 eV is reduced to 4.5 eV as experimentally observed. This reduction is most likely due to the hybridization of defect states with the conduction band. The golden shine of our highly reduced samples (see Fig.~\ref{Fig:beauty}) requires an absorption edge of about 2 eV. This feature can be explained by transition from the broadened midgap band to the conduction band.

The Fermi level sits inside a second defect state band in the vicinity of the conduction band. Due to the more than half filling of this band, we obtain $p$-type charge carriers. Currently, we are investigating our samples in more detail by optical spectroscopy in order to confirm more quantitatively our rough picture as concluded from the reported measurements

Takeuchi {\em et al}. have performed spectroscopic ellipsometry on hafnia thin films identifying an absorption peak at 4.5 eV counted from the upper edge of the valence band.\cite{Takeuchi:04} The oxygen deficiency was controlled by the oxidation time of hafnium metal thin films. The width of the defect band considerably broadens the more deficient the investigated films are. An overall decrease in optical band gap with decreasing annealing time in oxygen could be observed as well. This picture is qualitatively in agreement with our results.


\section{Summary}

In conclusion, we have performed oxygen engineering in hafnium oxide by reactive molecular beam epitaxy. The reproducible and stable introduction of oxygen vacancies allows the control of structural and electronic properties. Most notably, the band gap is reduced by more than one 1 eV, and the conductivity changes by several orders of magnitude. The dominant defect state is a single negative charged oxygen vacancy. At high concentrations of such defects, the electronic states overlap to form defect bands within the band gap explaining all experimentally observed physical properties of HfO$_{2-x}$. One upcoming application of controlled oxygen engineered hafnia thin films are resistive switching devices for non-volatile memory.


\section{Acknowledgements}

We acknowledge T.~Schr\"{o}der for the XPS measurements and helpful discussions, H.-J.~Kleebe, for obtaining HRTEM images, S.~Flege for SIMS measurements, and G.~Haindl for technical support.

This work was supported by DFG through Grant No. AL 560/13-1 and the LOEWE-Centre AdRIA.

\end{document}